\documentclass[prb,twocolumn]{revtex4}%

\usepackage{graphicx}
\usepackage{amsmath}
\usepackage{ulem}
\usepackage{color}

\newcommand{\be}{\begin{equation}}
\newcommand{\ee}{\end{equation}}
\newcommand{\bea}{\begin{eqnarray*}}
\newcommand{\eea}{\end{eqnarray*}}
\newcommand{\bean}{\begin{eqnarray}}
\newcommand{\eean}{\end{eqnarray}}

\begin{document}

\draft
\title
{\bf Heat diodes made of quantum dots embedded in nanowires
connected to metallic electrodes}

\author{ David M T Kuo}

\address{Department of Electrical Engineering and Department of Physics, National Central
University, Chungli, 320 Taiwan}



\date{\today}

\begin{abstract}
The quantum dot arrays (QDAs) embedded into inhomogeneous
nanowires connected to metallic electrodes show an electron heat
rectification effect, which is attributed to the thermal voltage
arising from a temperature bias and the QDA with a broken spatial
inversion symmetry. The staircase energy levels of QDAs can be
controlled to the resonant and off resonant transports for
electrons in the forward and backward temperature biases,
respectively. The effect of electron Coulomb interactions on the
rectification efficiency of heat diode is clarified by the case of
double QDs. We find that it is important to reduce phonon heat
currents for implementing a high efficient electron heat diode at
high temperature.
\end{abstract}

\maketitle

\textbf{1. Introduction}

Solid state heat diodes (HDs) are important for the applications
of energy harvesting, which transfers the wasted heats as useful
electrical powers.$^{1)}$ Many theoretical efforts have devoted to
design HDs.$^{2-14)}$ The rectification efficiency of HDs is
defined as $R=Q_{F}/|Q_B|$, where $Q_F$ and $Q_B$ are the heat
currents in the forward and backward temperature bias,
respectively. The ideal HDs will have vanishingly small $Q_B$
values, which indicate that heat currents in the backward
temperature bias are almost fully blockaded. So far, some systems
experimentally show the functionality of HDs, whereas their $R$
values are very small.$^{15-19)}$ Most recently $R=140$ was
reported in metal/superconductor junction system operated at
extremely low temperatures (below the boiling point of liquid
helium). Because it is infeasible to directly measure the heat
currents, $R=140$ is predicted by the theoretical model without
considering phonon heat currents.$^{20)}$ Meanwhile, BCS-type
superconductor junctions limit the operation of electron heat
diode in a small temperature range.

Unlike electron HDs of ref.[20], some designs have focused on
phonon HDs.$^{2-11)}$ In references[2-5] phonon HDs are designed
by considering exact one-dimensional systems. Such 1-D models
degrade the realistic applications of HDs.$^{6-11)}$ 3-D systems
are requested to give a promising application. Although
theoretical concepts about the phonon HDs designed in wherever
nonlinear interactions and broken inversion symmetry are
ubiquitous, currently available techniques have limited
sensitivities and are unable to unravel the interesting effect in
these phonon systems.$^{12)}$ Therefore, it is desirable to design
HDs which are feasible to be measured by thermoelectric technique.
Some 3-D semiconductor systems with vacuum layers are proposed to
design HDs.$^{13,14,17)}$ Here, we propose the HDs made of quantum
dot arrays (QDAs) with staircase energy levels embedded into
nanowires connected to metallic electrodes. Unlike phonon
HDs,$^{2-11)}$ our design can be applied to energy harvesting for
the applications of thermoelectric devices in the nonlinear
response regime.$^{21)}$

\textbf{2. Formalism}

To study the electron heat rectification of QDAs embedded into
nanowires connected to metallic electrodes shown in the inset of
Fig. 1(a), we start the system Hamiltonian given by an extended
Anderson model $H=H_0+H_{QD}$,$^{22)}$ where
\begin{eqnarray}
H_0& = &\sum_{k,\sigma} \epsilon_k
a^{\dagger}_{k,\sigma}a_{k,\sigma}+ \sum_{k,\sigma} \epsilon_k
b^{\dagger}_{k,\sigma}b_{k,\sigma}\\ \nonumber &+&\sum_{k,\sigma}
V^L_{k,L}d^{\dagger}_{L,\sigma}a_{k,\sigma}
+\sum_{k,\sigma}V^R_{k,R}d^{\dagger}_{R,\sigma}b_{k,\sigma}+c.c.
\end{eqnarray}
The first two terms of Eq.~(1) describe the free electron gas in
the left and right electrodes. $a^{\dagger}_{k,\sigma}$
($b^{\dagger}_{k,\sigma}$) creates  an electron of momentum $k$
and spin $\sigma$ with energy $\epsilon_k$ in the left (right)
electrode. $V^L_{k,L}$ ($V^R_{k,R}$) describes the coupling
between the left (right) QD system and left (right) electrode.
$d^{\dagger}_{L(R),\sigma}$ ($d_{L(R),\sigma}$) creates (destroys)
an electron in the left (right) QD.

\begin{eqnarray}
H_{QD}&=&\sum_{\ell,\sigma} E_{\ell} n_{\ell,\sigma}+ \sum_{\ell}
U_{\ell} n_{\ell,\sigma} n_{\ell,\bar\sigma}\\ \nonumber
&+&\frac{1}{2}\sum_{\ell,j,\sigma,\sigma'}
U_{\ell,j}n_{\ell,\sigma}n_{j,\sigma'}+\sum_{\ell \neq j}
t_{\ell,j} d^{\dagger}_{\ell,\sigma} d_{j,\sigma}+ c.c
\end{eqnarray}

where { $E_{\ell}$} is the spin-independent QD energy level, and
$n_{\ell,\sigma}=d^{\dagger}_{\ell,\sigma}d_{\ell,\sigma}$,
$U_{\ell}$ and $U_{\ell,j}$ describe the intradot and the nearest
interdot Coulomb interactions, respectively. $t_{\ell,j}$
describes the electron nearest neighbor hopping strength. Because
we consider the nanoscale semiconductor QDs, there is only one
energy level for each QD.

The electron and heat currents from left (right) electrode to the
QDA can be derived by using the Meir-Wingreen formula$^{23)}$. We
have
\begin{eqnarray}
J=\frac{2e}{\hbar}\int \frac{d\epsilon}{2\pi}{\cal
T}_{LR}(\epsilon)[f_{L}(\epsilon)-f_R(\epsilon)],
\end{eqnarray}
where
$f_{\alpha}(\epsilon)=1/\{\exp[(\epsilon-\mu_{\alpha})/k_BT_{\alpha}]+1\}$
denotes the Fermi distribution function for the $\alpha$-th
electrode, where $\mu_\alpha$  and $T_{\alpha}$ are the chemical
potential and the temperature of the $\alpha$ electrode. $e$,
$\hbar$, and $k_B$ denote the electron charge, the Planck's
constant, and the Boltzmann constant, respectively. ${\cal
T}_{LR}(\epsilon)$ denotes the transmission coefficient of QDA
embedded in a nanowire connected to electrodes. The heat current
for electrons leaving from the left (right) electrode is given by

\begin{equation}
Q_{e,L(R)}=\frac{\pm 2}{\hbar}\int \frac{d\epsilon}{2\pi}{\cal
T}_{LR}(\epsilon)(\epsilon-\mu_{L(R)})[f_{L}(\epsilon)-f_R(\epsilon)].
\end{equation}
We note that $Q_L+Q_R=-(\mu_L-\mu_R)\times J/e$ gives a Joule
heating.$^{24)}$ To discuss the electron heat rectification, we
consider the condition of open circuit ($J=0$) under a temperature
bias $\Delta T=T_L-T_R$, where $T_L=T+\Delta T/2$ and
$T_R=T-\Delta T/2$. $T$ denotes the averaged temperature of
junction system. Due to the Seebeck effect, the thermal voltage
$V_{th}$ arising from $\Delta T$ will balance the diffusing
electrons from the hot electrode to the cold electrode. Meanwhile,
the energy level $E_{\ell}$ of each QD will be shifted due to
$V_{th}$. As a consequence, ${\cal T}_{LR}(\epsilon)$ will depend
on $V_{th}$.

Because it is difficult to fully blockade phonon heat currents
$Q_{ph}$ in semiconductor junction systems,$^{1)}$ $Q_{ph}$ is
considered by an empirical formula in Ref. [25], which allow us to
avoid the detailed phonon dispersion structures. It's expression
is given by
\begin{equation} Q_{ph}(T)= \frac{F_s}{\hbar} \int \frac{d\omega}{2\pi} {\cal T}(\omega)_{ph}
(\hbar^2 \omega)[n_{L}(\omega)-n_R(\omega)], \label{phC}
\end{equation}
where $\omega$ and ${\cal T}_{ph}(\omega)$ are the phonon
frequency and throughput function, respectively.
$n_{L(R)}=1/(exp(\omega/T_{L(R)})-1)$. We assume that phonon
temperature is the same as electron temperature. Dimensionless
factor $F_s$ is used to describe the phonon scattering resulting
from the interface scattering of QDs embedded into
nanowire.$^{26,27)}$ The value of $F_s=0.1$ is used through this
article. $F_s=0.1$ is determined by the theoretical results of
ref.[27], where authors calculated the phonon thermal conductance
of silicon/germanium QD nanowires. The expression of throughput
function is given in Ref. [25], which reads
\begin{equation}
{\cal T}_{ph}(\omega)=\frac{N_{ph,1}(\omega)}{1+
L/\ell_0(\omega)}+\frac{N_{ph,2}(\omega)}{1+L/D}
\end{equation}
with the frequency-dependent mean free path $\ell_0(\omega)$ given by
\begin{equation}
\frac{1}{\ell_0(\omega)}=B\frac{\delta^2}{D^3}(\frac{\omega}{\omega_D})^2 N_{ph}(\omega),
\end{equation}
where
$N_{ph}(\omega)=4+A(\frac{D}{a})^2(\frac{\omega}{\omega_D})^2$
(for $\omega < \omega_D$) denotes the number of phonon modes. The
dimensionless parameters are chosen to be $A=2.17$ and
$B=1.2$.[25] Notation $a$ denotes the lattice constant of
nanowire. $N_{ph,1}(\omega)=N_{ph}(min(\omega,v_s/\delta))$ and
$N_{ph,2}(\omega)=N_{ph}(\omega)-N_{ph,1}(\omega)$. $v_s$ is the
sound velocity of nanowire and $\delta$ describes the width of the
disorder surface of nanowires.$^{24)}$ In Eq.~(5), one essentially
replaces the frequency-dependent mean free path $\ell_0(\omega)$
by a diameter of nanowire ($D$) for the high-frequency modes
($\omega > v_s/\delta$). The phonon thermal conductance calculated
by Eq. (5) under small temperature bias ($\Delta T$) and $F_s=1$
well explains the experimental results of silicon
nanowires.$^{25,28)}$

\textbf{3. Results and discussion }

Based on Eq. (3), we numerically solve the thermal voltage
($V_{th}$) self-consistently by considering the condition of
$J=0$. Substituting $V_{th}$ into  Eq. (4), we can calculate the
electron heat current of $Q_{e}$. For the inhomogeneous nanowire
shown in the inset of Fig. 1(a), $E_n$ will depend on the QD
location. Here, we assume QDA with staircase energy levels shown
in the inset of Fig. 1(b) in which $E_1=E_R$, $E_2=E_R+\Delta E$,
and $E_n=E_R+(n-1)\Delta E$, where $\Delta E$ is a uniform energy
level separation. Due to $V_{th}$, $E_n$ will be replaced by
$\epsilon_n=E_n+\eta_D eV_{th}$, where factor $\eta_D$ is
determined by the QD location, shape and material dielectric
constant.$^{29)}$ For the simplicity, $\eta_D$ is determined by QD
location $\eta_D=(z-L/2)/L$, where $L$ denotes the length of QD
nanowire. We have the QD location $z=n\times L_p$, where $L_p$ is
the lattice constant of QD superlattice.$^{27)}$ The parameters of
QD number $N=25$, $L_p=5~nm$ and $L=127~nm$ are adopted in our
calculations. The expression of ${\cal T}_{LR}(\epsilon)$ for N=25
in terms of a single particle retarded Green's functions can be
found in our previous work.$^{30)}$ It is a prohibited challenge
to include all correlation functions arising from electron Coulomb
interactions for $N=25$.$^{31,32)}$ The affect of electron Coulomb
interactions on the electron transport of double QDs is clarified
in Fig. 5 in which we reveal why electron Coulomb interactions can
be neglected in this study of $N=25$. In Fig. 1(a) we show the
heat currents as a function of averaged temperature for the
forward ($T_L > T_R$) and backward ($T_R > T_L$) temperature
biases ($|\Delta T=24 K|$) at $\Delta E=0.1~meV$ and
$E_1=E_R=E_F+4~meV$. The tunneling rate between the left (right)
QD and left (right) electrode is given by
$\Gamma_{L(R)}(\epsilon)=\sum_{k}|V_{k,L(R)}|^2\delta(\epsilon-\epsilon_k)$.
In the wide band limit of electrodes, the energy-dependent
$\Gamma_{L(R)}(\epsilon)$ can be ignored. Here we have considered
$\Gamma_L=\Gamma_R=\Gamma=8~meV$. $Q_{e,F}$ is larger than
$|Q_{e,B}|$ in the whole temperature range. The heat rectification
effect is observed in Fig. 1(b) by  $R=Q_{e,F}/|Q_{e,B}|\neq 1$.
The maximum $R$ reaches 1.8 at temperature $T=38K$ for $\Delta
T=24~K$. $R$ decreases with increasing averaged temperatures at a
fixed temperature bias.

\begin{figure}[h]
\centering
\includegraphics[angle=-90,scale=0.3]{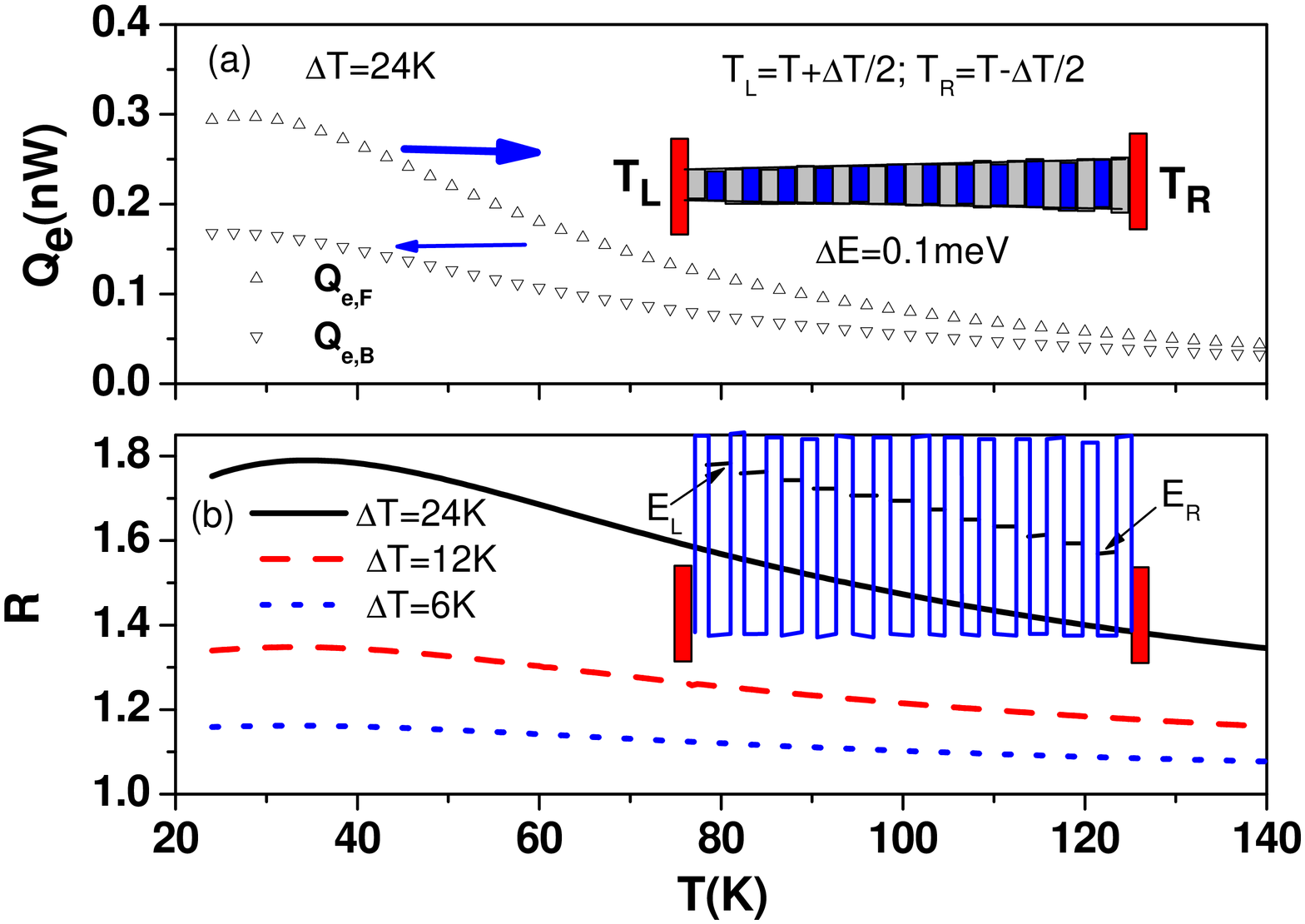}
\caption{  (a) Heat current $Q_e$, and (b)  rectification
efficiency $R$ as a function of averaged temperature in the
absence of electron Coulomb interactions and phonon heat currents.
We have adopted electron physical parameters $t_{c}=4~meV$,
$\Delta E=0.1~meV$ and $\Gamma=8~meV$.}
\end{figure}

To reveal the mechanism of electron heat rectification of Fig. 1,
we plot the thermal voltages in Fig. 2. Note that $V_{th}$ are
negative and positive values in the forward and backward
temperature biases, respectively. Nonlinear Seebeck coefficients
($V_{th}/\Delta T$) are always negative values, which illustrate
that electrons of the electrodes mainly diffuse through the energy
levels above $E_F$.  The thermal voltage not only changes the
chemical potentials of electrodes ($\mu_{L(R)}=E_F\pm eV_{th}/2$)
to against the diffusing electrons from the hot side to the cold
side, but also controls electrons of the electrodes tunneling
through resonant and off-resonant channels. In the forward
temperature bias the staircase energy levels of QDAs is tuned to
quasi resonant channels (see the inset of Fig. 2(a)). On the other
hand, QDA energy levels are in the off-resonant channels for
backward temperature bias (see the inset of Fig. 2(b)). As a
consequence, electron heat current in the $T_L > T_R $ is larger
than that in $T_L < T_R $. When QDA energy levels are resonant at
$E_{res}$ resulting from a specify $V_{th,res}$, QDA forms a
miniband, which is described by the transmission coefficient

\begin{equation}
{\cal T}_{LR}(\epsilon)=\Pi_{n=1}^{N\ge 2}\frac{\Gamma_L \Gamma_R
(t^2_c)^{N-1}
}{(\epsilon-(E_{res}-2t_c~cos(\frac{n\pi}{N+1})))^2+\Gamma^2_n}.
\end{equation}
$\Gamma_n$ depends on QD number. For example, $N=3$, we have
$\Gamma_1=\Gamma/4$, $\Gamma_2=\Gamma/2$ and $\Gamma_3=\Gamma/4$.
$N=4$, $\Gamma_1=\Gamma/8$, $\Gamma_2=3\Gamma/8$,
$\Gamma_3=3\Gamma/8$ and $\Gamma_4=\Gamma/8$. $N=5$,
$\Gamma_1=\Gamma/12$, $\Gamma_2=\Gamma/4$, $\Gamma_3=\Gamma/3$,
$\Gamma_4=\Gamma/4$ and $\Gamma_5=\Gamma/12$. $\Gamma_n$ satisfies
$\sum_{n=1}^N\Gamma_n=\Gamma$. Eq. (8) clearly illustrates the
miniband formed by 25 resonant channels with Lorenz-shape of QD
nanowire. When $V_{th}$ is larger than $V_{th,res}$, this miniband
will be lifted.

\begin{figure}[h]
\centering
\includegraphics[angle=-90,scale=0.3]{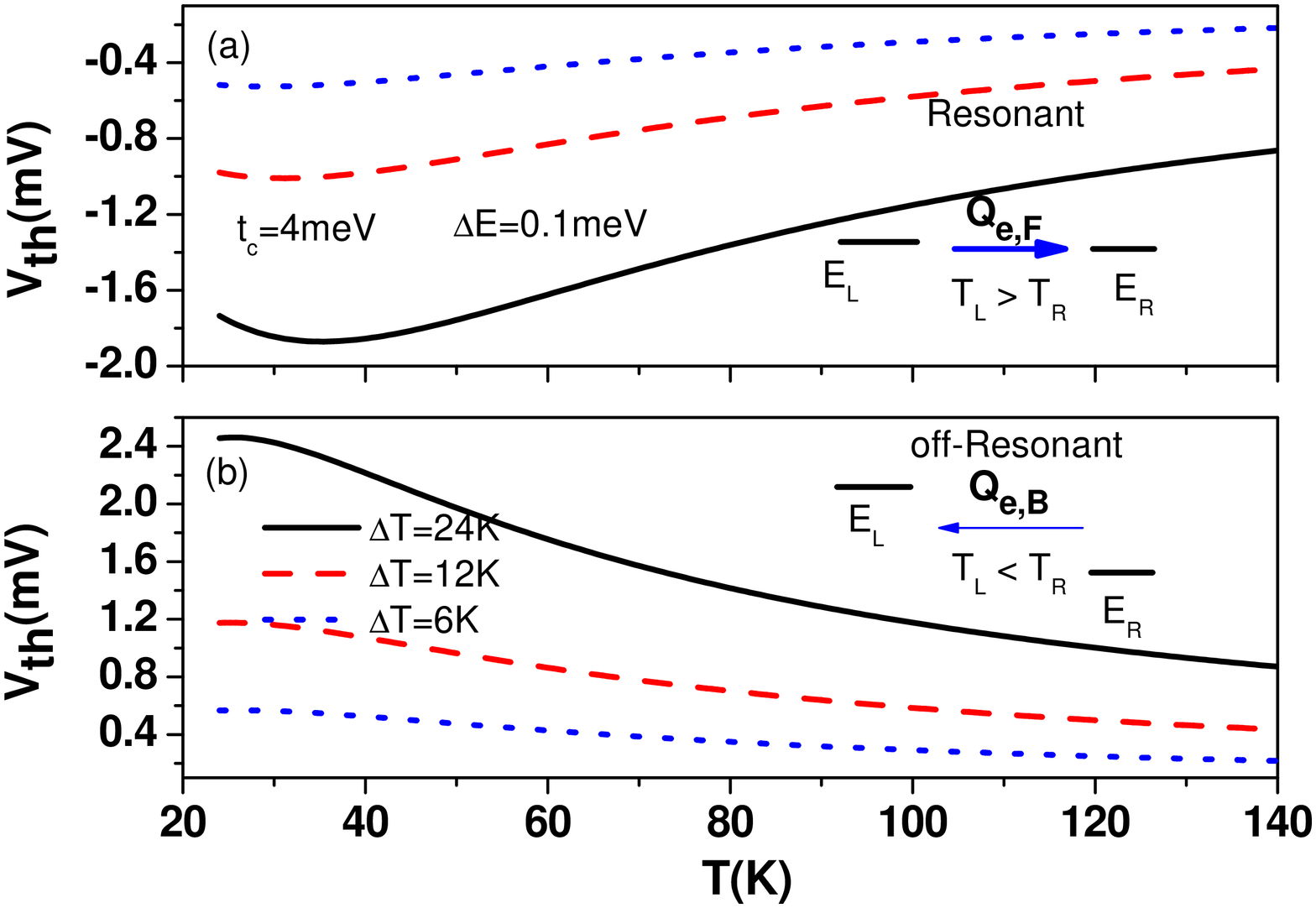}
\caption{ Thermal voltages $V_{th}$  as a function of $T$ for
different temperature biases at $\Delta E=0.1~meV$. Diagrams (a)
and (b) correspond to forward and backward temperature biases,
respectively. Other physical parameters are the same as those of
Fig. 1.}
\end{figure}

In addition to $V_{th}$, the design of staircase energy levels of
QDA with broken spatial inversion symmetry also plays a remarkable
role to observe the behavior of electron heat rectification. We
calculate $Q_e$ and $R$ as a function of $\Delta E$ for different
tunneling rates in Fig. 3. When $\Delta E=0$
($E_n=E_R=E_F+4~meV$), we see $Q_{e,F}=Q_{e,B}$ and $R=1$. This
indicates the vanish of rectification effect when QDA energy
levels have a spatial inversion symmetry. $Q_{e,F}$ and $Q_{e,B}$
decrease with increasing $\Delta E$, whereas $Q_{e,B}$ decreases
quickly. As a consequence, $R$ is highly enhanced. As for the
variation of tunneling rates, electron heat currents is reduced
with decreasing $\Gamma$. Nevertheless, $R$ is not sensitive to
the variation of $\Gamma$.

\begin{figure}[h]
\centering
\includegraphics[angle=-90,scale=0.3]{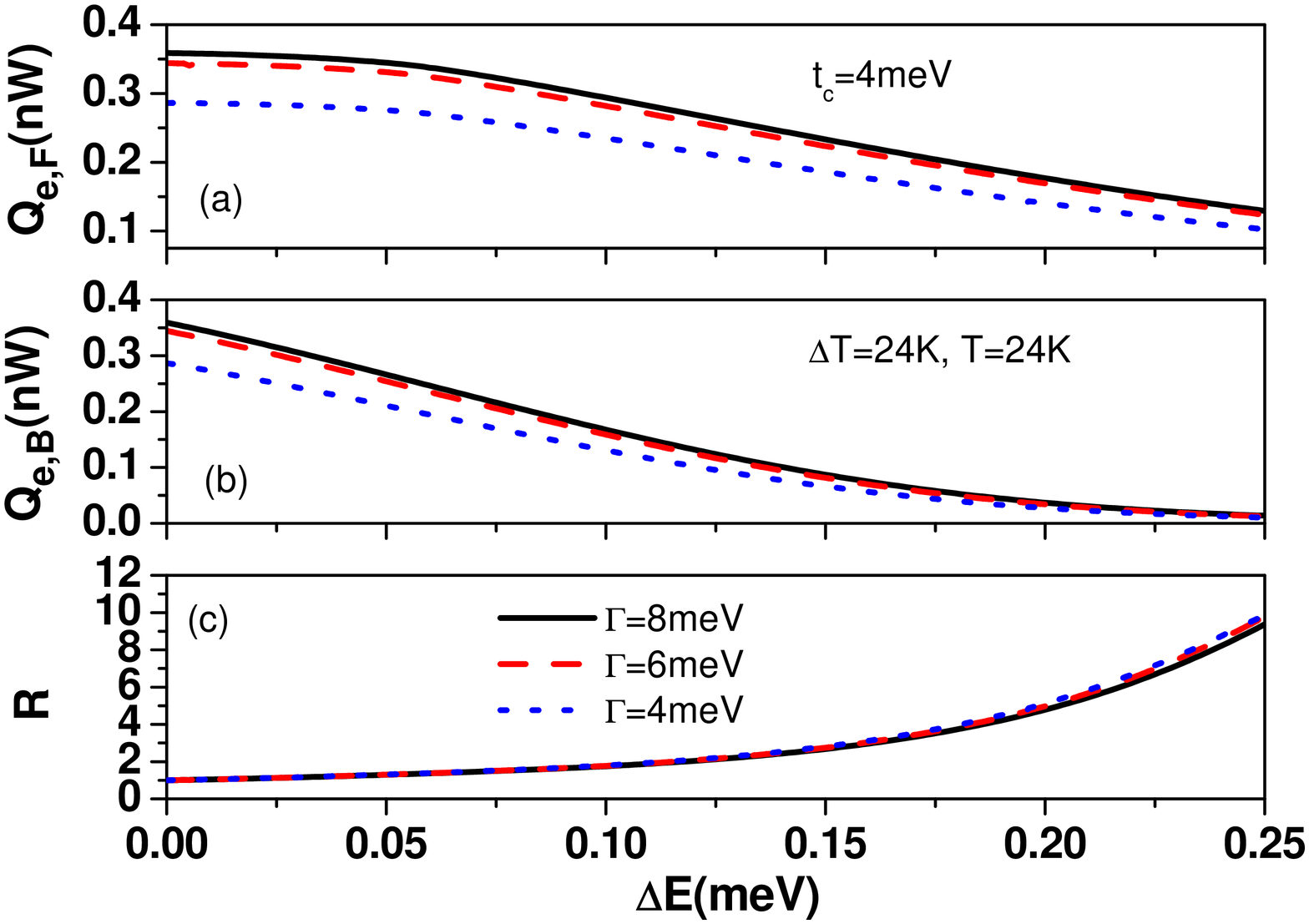}
\caption{(a) $Q_{e,F}$, (b) $Q_{e,B}$ and (c) $R$ as a function of
$\Delta E$ for different $\Gamma$ values at $t_c=4~meV$, $T=24~K$
and $\Delta T=24~K$.}
\end{figure}

For a heat diode, one has to examine the behavior of heat current
as a function of temperature bias.$^{2-14}$ We plot $Q_e$ as a
function of $\Delta T$ for different $E_R$ values at a averaged
temperature ($T=24~K$) in Fig. 4(a). Electron heat currents show a
manifested rectification behavior since $Q_{e,F}$ is much larger
than $Q_{e,B}$. However, $Q_e$ is suppressed with increasing
$E_R$. When QD energy levels are tuned away from $E_F$, electron
population decreases and then $Q_e$ is reduced. In addition, the
negative differential thermal conductance (NDTC) feature, which
$Q_{e,B}$ decreases with increasing $\Delta T$, exists in the
backward temperature bias regime. In Fig. 4(b) $R$ values are
larger than ten for dotted and dashed curves in the regime of
$\Delta T > 20~K$. Although $R$ values are smaller than those
reported in ref. [20], the operation averaged temperature is much
higher than that of ref.[20].

\begin{figure}[h]
\centering
\includegraphics[angle=-90,scale=0.3]{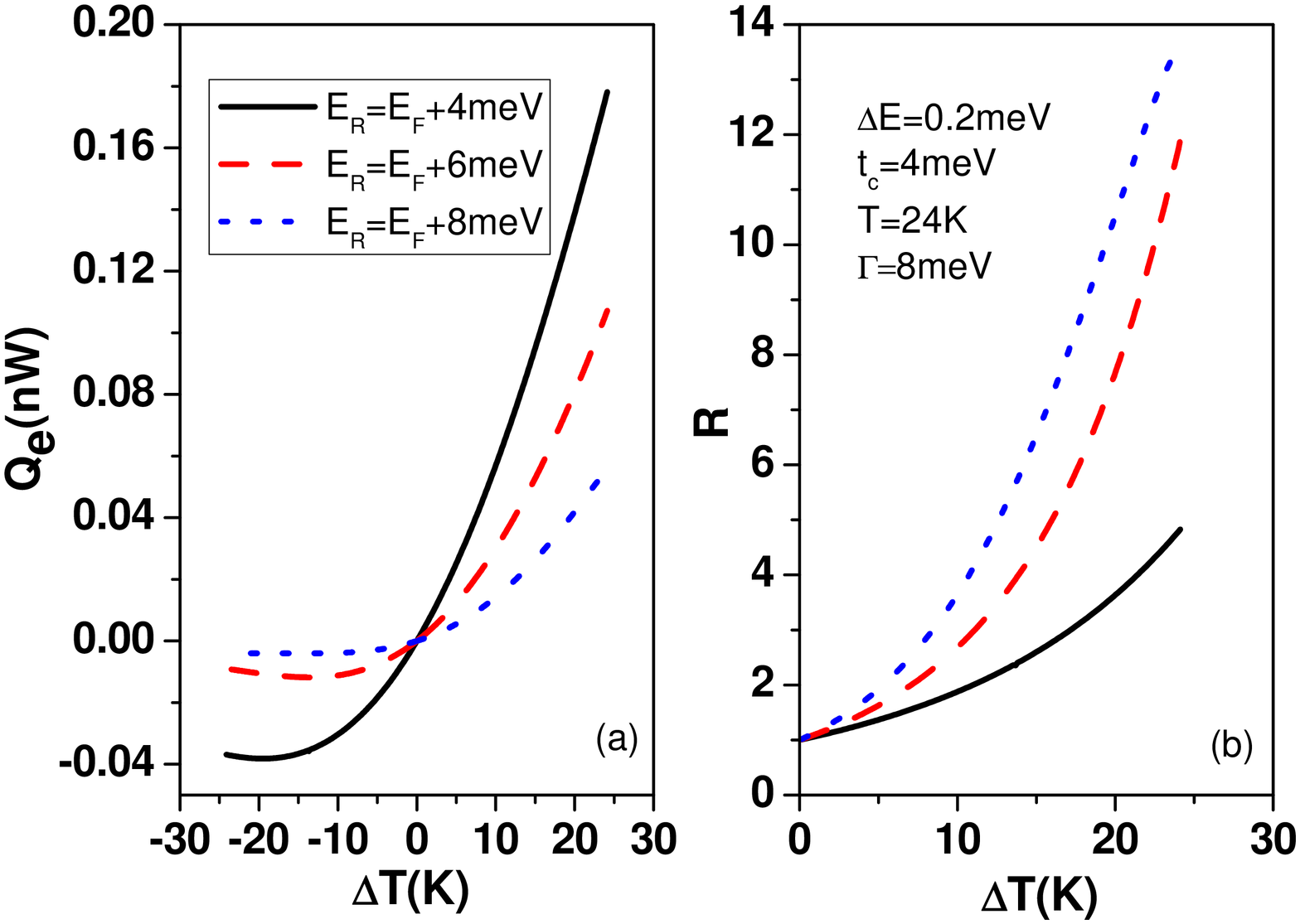}
\caption{(a)Electron heat current $Q_e$ and (b) rectification
efficiency $R$ as a function of temperature bias ($\Delta T$) for
different $E_R$ values at $\Delta E=0.2meV$, $t_c=4~meV$,
$\Gamma=8~meV$ and $T=24~K$.}
\end{figure}

In the previous figures, electron Coulomb interactions are not
included in calculations. We have to examine the effect of
electron Coulomb interactions on the electron heat rectification.
To clarify the effect of electron intradot and interdot Coulomb
interactions on $R$, the electron heat currents and $R$ values of
double QDs (DQDs) as a function of temperature bias for different
electron Coulomb interactions are plotted in Fig. 5. The
expression of ${\cal T}_{LR}(\epsilon)$ for DQDs can be found in
ref [31,28]. In Fig. 5(a) electron heat currents are suppressed in
the presence of electron Coulomb interactions. Such an effect is
more significant at large $\Delta T$ due to the increase of
electron population. Because $E_R=E_F+4~meV$ and $E_L=E_F+8~meV$
are above $E_F$, we can consider the approximated expression of
${\cal T}_{LR}(\epsilon)$ as
\begin{equation}
{\cal T}_{LR}(\epsilon)\approx\frac{4\Gamma_L\Gamma_R
t^2_{c}P_{1}}{|(\epsilon-\epsilon_L+i\Gamma_L)
(\epsilon-\epsilon_R+i\Gamma_R)-t^2_{c}|^2},
\end{equation}
where $\epsilon_L=E_L+\eta_D eV_{th}$ and $\epsilon_R=E_R-\eta_D
eV_{th}$. $P_1$ is the probability weight of double QDs with empty
state, which is determined by the one particle occupation numbers
and other correlation functions.$^{28,31}$ Other resonant levels
arising from intradot and interdot Cooulomb interactions are very
far away from $E_F$ can be ignored in Eq. (9). If one turns off
electron Coulomb interactions, $P_1=1$. From the results of Fig.
5(a), $P_1$ is just slightly reduced. This indicates that the
calculations of figures (1)-(4) will be slightly changed as long
as $U_{\ell,j}/(k_B\Delta T) \gg 1$, which is the regime we are
interested in. Fig. 5(b) show the rectification efficiency for
different electron Coulomb interaction configurations. Although
$Q_e$ is suppressed in the presence of $U_{\ell}$ and
$U_{\ell,j}$, $R$ is enhanced. This enhancement due to electron
Coulomb interactions will disappear in the presence of phonon heat
currents.
\begin{figure}[h]
\centering
\includegraphics[angle=-90,scale=0.3]{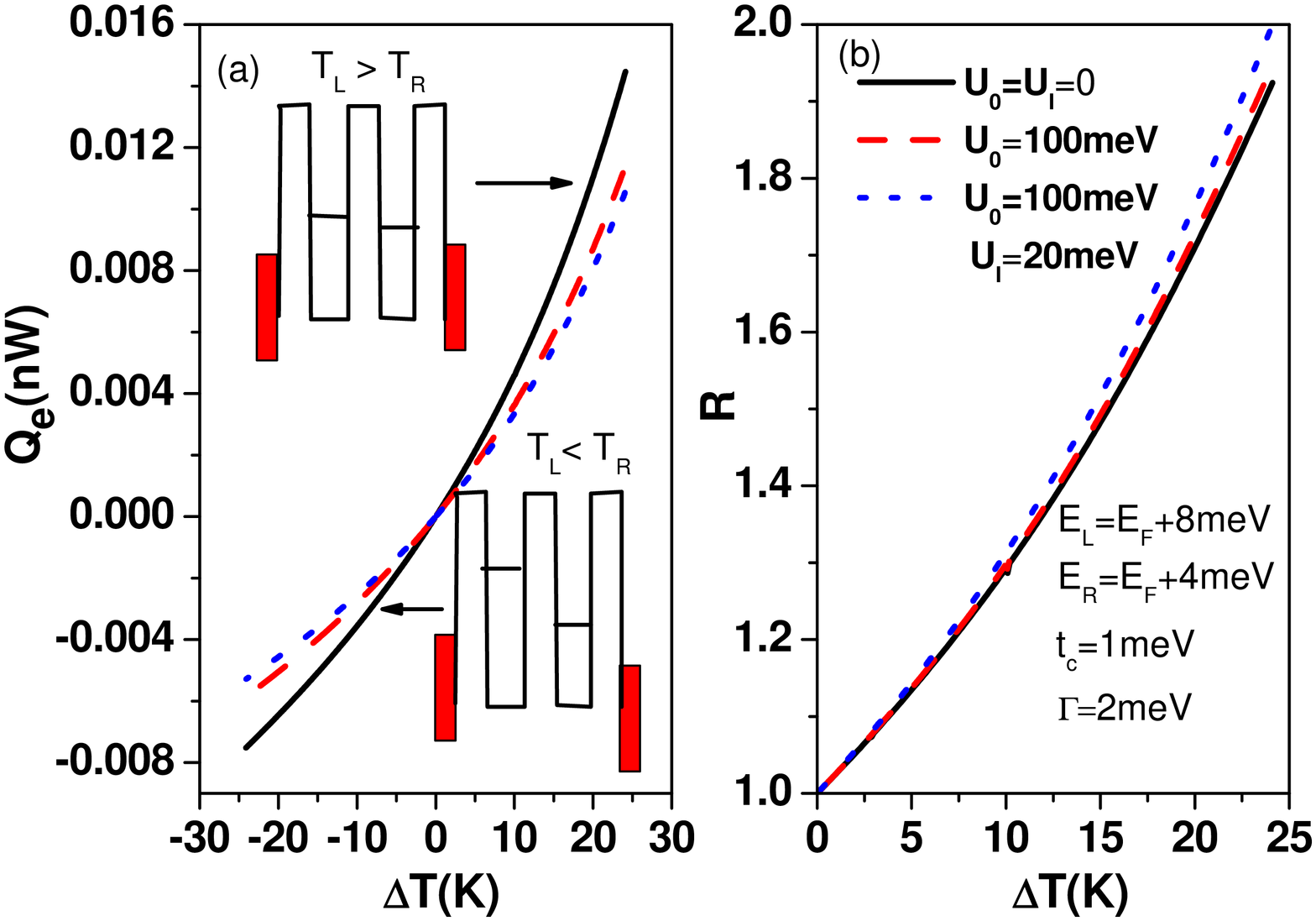}
\caption{(a) Electron heat current $Q_e$ and (b) rectification
efficiency $R$ as a function of temperature bias for different
electron interaction configurations in the case of double QDs. We
have $E_R=E_F+4~meV$, $E_L=E_F+8~meV$, $t_c=1~meV$ $\Gamma=2~meV$
and $\eta_D=0.3$.}
\end{figure}

Although a vacuum layer allow high efficient electron HDs operated
at room temperature, it increases the technique
complication.$^{14)}$ Finally, we clarify how phonon heat currents
to influence the electron heat rectification. Based on Eq. (5), we
calculate $Q_{ph}$ as a function of averaged temperature for
different diameters of silicon/germanium QD nanowires at a fixed
temperature bias $\Delta T=24~K$ and disorder surface width
$\delta=2~nm$ in Fig. 6(a). The behavior of $Q_{ph}$ can be
understood by $Q_{ph}\approx \kappa_{ph}\Delta T$, where
$\kappa_{ph}$ is phonon thermal conductance. Therefore, Fig. 6(a)
reveals that $\kappa_{ph}$ increases with increasing averaged
temperature.$^{25,28)}$ Such a temperature behavior of
$\kappa_{ph}$ between $20~K$ and $140$ is not limited in the
silicon/germanium QD systems, but the typical feature of nanowires
with surface disorder effects.$^{1,25)}$ Phonon heat currents
decrease with decreasing $D$. The $Q_{ph}$ values are smaller than
$Q_{e}$ as $T < 40 K$, but their magnitudes are comparable to
$Q_e$. To evaluate $R$, obviously we have to include the phonon
heat currents. In Fig. 6(b) we calculate
$R=Q_{e,F}+Q_{ph}/(|Q_{e,B}|+Q_{ph})$ for different $D$ values at
$\Delta E=0.2~meV$. In the absence of $Q_{ph}$, we observe that
$R$ is larger than 3 when $T$ is below $60 K$. Once $Q_{ph}$ is
included, $R$ is suppressed seriously. The results of Fig. 6(b)
imply that it is crucial to have  phonon glass
materials$^{33-35)}$ or some novel ideas (for example phonon
localization) to fully blockade $Q_{ph}$ for high efficient
electron heat diodes. Because silicon and germanium are not polar
semiconductors, the electron phonon interactions (EPIs) are
vanishingly small in a finite length nanowire. Therefore, we can
ignore the effect of EPIs on electron and phonon heat currents.
\begin{figure}[h]
\centering
\includegraphics[angle=-90,scale=0.3]{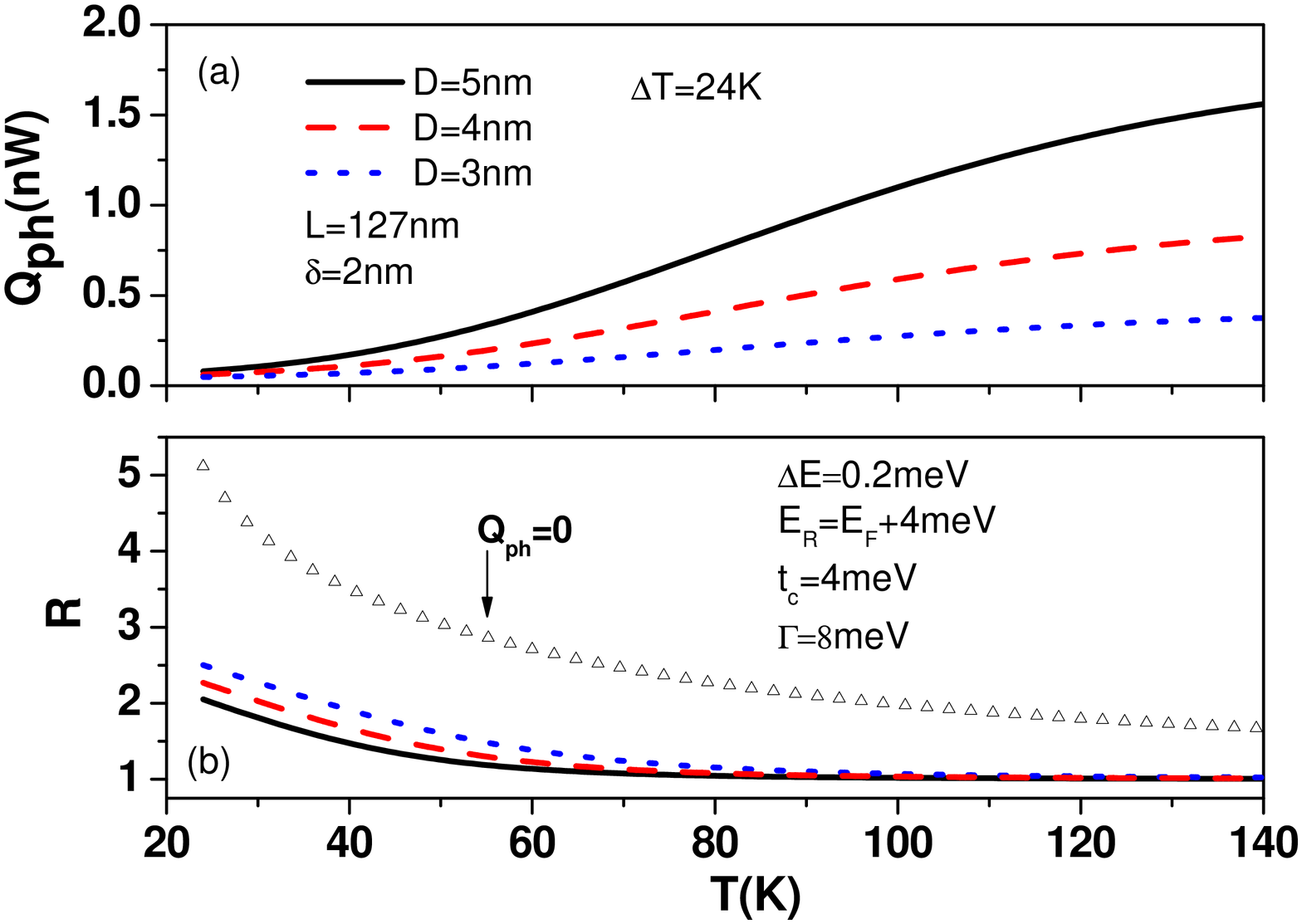}
\caption{(a)Phonon heat current $Q_{ph}$ and (b) rectification
efficiency $R$ as a function of averaged temperature  for
different diameters of nanowires with length $L=127~nm$, and
surface disorder width $\delta=2~nm$. Other physical parameters
are the same as those of silicon semiconductors.}
\end{figure}

\textbf{4. Conclusion}

We have theoretically investigated the electron heat rectification
of QDA embedded into a nanowire. The staircase energy levels with
a broken spatial inversion symmetry and thermal voltage resulting
from temperature biases play a remarkable role to observe electron
heat rectification effect. The staircase energy levels of QD
nanowire may be achieved by considering inhomogeneous QD nanowire
shown in Fig. 1(a). Recently, we have demonstrated that the
electron conductance and electron thermal conductance of DQDs can
be highly enhanced by increase of level degeneracy.$^{28)}$ The
rectification efficiency of electron HDs may be enhanced based on
level degeneracy.



{\bf Acknowledgments}
This work was supported by the Ministry of Science and Technology
of Taiwan under Contract No. MOST 103-2112-M-008-009-MY3
\mbox{}\\
E-mail address: mtkuo@ee.ncu.edu.tw\\

\setcounter{section}{0}

\renewcommand{\theequation}{\mbox{A.\arabic{equation}}} 
\setcounter{equation}{0} 

\mbox{}\\


\end{document}